\documentclass[6pt,a4paper,twocolumn]{article}
\usepackage[utf8]{inputenc}
\usepackage[T1]{fontenc}
\usepackage{mathptmx}
\usepackage[pdftex]{graphicx}
\usepackage[pdftex,linkcolor=black,pdfborder={0 0 0}]{hyperref}
\usepackage{calc}
\usepackage{enumitem}

\usepackage[a4paper, lmargin=0.08\paperwidth, rmargin=0.08\paperwidth, tmargin=0.08\paperheight, bmargin=0.08\paperheight]{geometry}

\usepackage[all]{nowidow}
\usepackage[protrusion=true,expansion=true]{microtype}

\usepackage{upgreek}
\usepackage{amsmath}
\usepackage{color}
\usepackage{amsthm}
\usepackage{xspace}
\usepackage[mathscr,mathcal]{euscript}
\usepackage{amsfonts}
\usepackage{hyperref}
\usepackage{complexity}

\newtheorem{theorem}{Theorem}

\newcommand{\ie}{i.e.,\xspace~}

\newcommand{\eg}{e.g.,\xspace~}

\newcommand{\stops}{\mathcal{S}}
\newcommand{\trips}{\mathcal{T}}
\newcommand{\aTrip}{T}
\newcommand{\aRoute}{R}
\newcommand{\cliqueCover}{\mathcal{C}}

\newcommand{\absoluteVal}[1]{\left\vert #1 \right\vert}

\newcommand{\stopSeq}[1]{\stops\left( #1 \right)}

\author{Steil, Patrick}
\title{Optimal FIFO grouping in public transit networks}
\makeatletter
\hypersetup{
    pdfsubject = {Finding an optimal grouping of trips into routes such that no two trips of the same line overtake each other is a problem in~$\P$},
    pdftitle = {\@title},
    pdfauthor = {\@author}
}

\usepackage{fancyhdr}
\begin{document}
	\vspace{-\baselineskip}
	\section{Introduction}
	\vspace{-.8\baselineskip}
	This technical report is about grouping public transit vehicles (\textit{trips}) into a set of routes (or, as they are more commonly known: \textit{lines}) so that two vehicles of the same route do not overtake each other. We say that such a set of routes satisfies the FIFO property. A natural question is: Given a set of trips, find a minimal FIFO grouping into routes. This is especially interesting for certain route planning algorithms since a better (\textit{smaller}) route grouping leads to a better runtime. This contribution is structured as follows: First, all necessary details are formalised and defined, and then the algorithmic complexity of this problem is explained and proven.
	\vspace{-1.2\baselineskip}
	\section{Definitions}
	\vspace{-.8\baselineskip}
	We now introduce notations and definitions. 
	A timetable consists of a set of stations~$\stops$, \ie places where passengers can hop on and off vehicles, and a set of trips~$\trips$, \ie vehicles that travel through the network. 
	A trip~$\aTrip \in \trips$ consists of a sequence of (chronological) events, where the~$i$'th event~$\aTrip[i] = \uptau^i$ represents the arrival and departure of the vehicle at the~$i$'th stop~$\uptau^i_{\mathrm{stop}} \in \stops$ along its stop sequence~$\stopSeq{\aTrip}$. 
	The departure time of~$\uptau$ is indicated by~$\uptau_{\mathrm{dep}}$, the arrival time by~$\uptau_{\mathrm{arr}}$. 
	We require~$\uptau^{i}_{\mathrm{arr}} \leq \uptau^{i}_{\mathrm{dep}} \leq \uptau^{i+1}_{\mathrm{arr}} \leq \uptau^{i+1}_{\mathrm{dep}}\;\forall i \in \left[\absoluteVal{\aTrip}-1\right]$\footnote{$\left[n\right] \equiv \left[1, n\right] \cap \mathbb{N}$}. 
	Given two trips~$A \neq B \in \trips,\;\stopSeq{A} = \stopSeq{B}$, we define~$A \preceq B$ as ``$A$ being earlier than~$B$'' if the following conditions hold $\forall i \in \left[\absoluteVal{A}\right]$:
	\begin{align}
		\label{eq:1}
		& A[i]_{\mathrm{stop}} = B[i]_{\mathrm{stop}}\\
		\label{eq:2}
		& \left(A[i]_{\mathrm{arr}} \leq B[i]_{\mathrm{arr}}\right) \;\text{and}\; \left(A[i]_{\mathrm{dep}} \leq B[i]_{\mathrm{dep}}\right)
	\end{align}
	If~$\exists j \in \left[\absoluteVal{A}\right]: \left(A[j]_{\mathrm{arr}} < B[j]_{\mathrm{arr}}\right) \vee \left(A[j]_{\mathrm{dep}} < B[j]_{\mathrm{dep}}\right)$, we write~$A \prec B$. 
	We say~$A$ and~$B$ do not ``overtake'' each other if either~$A \preceq B$ or vice versa. 
	A route~$\aRoute = \left\{\aTrip_1, \aTrip_2, \dots\right\}$ is a set of trips, such that one can order all trips~$\aTrip_1 \preceq \aTrip_2 \preceq \dots$, \ie no trip overtakes another trip on this route~$\aRoute$.
	A directed \textit{graph}~$G = \left(V, E\right)$ is a tuple of vertices~$V$ and edges~$E \subseteq V \times V$.
	A \textit{comparability} graph is a directed graph~$G = \left(V, E\right)$, with a transitive orientation $F \subseteq E$. 
	An orientation~$F$ is defined as~$\left(v, w\right) \in F \iff \left(w, v\right) \notin F$ such that every edge~$e \in E$ is ``oriented'' by~$F$. 
	For the following definitions, let $G = \left(V, E\right)$ be a graph.
	We call a set of vertices~$C \subseteq V$ a \textit{clique}, if all vertices are pairwise connected. 
	The \textit{clique cover} problem asks for a partition~$\cliqueCover$ of~$V$ into cliques. 
	Finding a minimal clique cover for a graph is~$\NP$-hard~\cite{inproceedings}.
	\vspace{-\baselineskip}
	\section{Problem Statement \& Proof}
	\vspace{-.8\baselineskip}
	One trivial and valid FIFO grouping is to allow a new route for each trip. 
	As mentioned in the introduction, it is necessary, from an algorithmic point of view, to minimise the number of FIFO routes. 
	Nevertheless, how hard is it to find an optimal FIFO grouping? 
	We prove this problem is solvable in~$\P$, meaning a polynomial time algorithm exists to find such an optimal FIFO grouping.
	\begin{theorem}
		\label{th:one}
		Given a set of trips~$\trips$, finding a minimal FIFO route grouping such that no two trips of the same route overtake each other is solvable in polynomial time.
	\end{theorem}
	\vspace{-.5\baselineskip}
	\begin{proof}
		We will find such a FIFO route grouping by transforming the problem into multiple comparability graphs and finding a smallest clique cover~$\cliqueCover$ for each of these graphs. 
		Each clique~$C \in \cliqueCover$ corresponds to a route~$\aRoute$. Since there exists a polynomial time algorithm which computes~$\cliqueCover$ given a comparability graph~\cite{HOANG1994133}, Theorem~\ref{th:one} follows. 
		For every distinct stop sequence~$\sigma = \left<s_1, s_2, \dots, s_k\right>$ which occurs in~$\trips$ (meaning~$\exists \aTrip \in \trips: \stopSeq{\aTrip} = \sigma$), we create a graph~$G_{\sigma} = \left(\left\{\aTrip \in \trips \;\vert\; \stopSeq{\aTrip} = \sigma\right\}, F_{\sigma}\right)$. $F_{\sigma}$ is defined by the following (with~$\aTrip \neq \aTrip' \in G_{\sigma}$):
		\begin{equation}
			\left(\aTrip, \aTrip'\right) \in F_{\sigma} \iff \aTrip \preceq \aTrip'
		\end{equation}
		We need to show that~$F_{\sigma}$ is a transitive orientation. 
		Let~$A \neq B \neq C \in G_{\sigma}$ and $\left(A, B\right), \left(B, C\right) \in F_{\sigma}$. 
		We show that~$\left(A, C\right) \in F_{\sigma}$, \ie both equations~(\ref{eq:1}) and~(\ref{eq:2}) hold. 
		Note that~(\ref{eq:1}) follows by construction of~$G_{\sigma}$. As for~(2): 
		We know, that~$\forall i \in \left[\absoluteVal{A}\right]$
		\begin{align*}
			& \left(A[i]_{\mathrm{arr}} \leq B[i]_{\mathrm{arr}}\right) \;\text{and}\; \left(A[i]_{\mathrm{dep}} \leq B[i]_{\mathrm{dep}}\right) \\
			& \left(B[i]_{\mathrm{arr}} \leq C[i]_{\mathrm{arr}}\right) \;\text{and}\; \left(B[i]_{\mathrm{dep}} \leq C[i]_{\mathrm{dep}}\right)
		\end{align*}
		hold. 
		It follows:
		\vspace{-.6\baselineskip}
		\begin{align*}
			\Rightarrow & \left(A[i]_{\mathrm{arr}} \leq C[i]_{\mathrm{arr}}\right) \;\text{and}\; \left(A[i]_{\mathrm{dep}} \leq C[i]_{\mathrm{dep}}\right) \\
			\Rightarrow & A \preceq C \iff \left(A, C\right) \in F_{\sigma}
		\end{align*}
		\vspace{-.6\baselineskip}
	\end{proof}
    \vspace{-\baselineskip}
    \paragraph*{Note} Preliminary results on real-life datasets (GTFS feeds from \eg Germany, Switzerland, \dots) show that a greedy approach finds the same number of FIFO routes as the optimal algorithm. Both algorithms are implemented here \url{https://github.com/TransitRouting/Arc-FlagTB}. This was to be expected since vehicles of the same route do not overtake each other or relatively rarely in reality.
	\vspace{-\baselineskip}
	\paragraph*{Acknowledgement}
	I thank Jonas Sauer, with whom I discussed this problem and the solution.
	\vspace{-1.2\baselineskip}
	\bibliographystyle{plainurl}
	\bibliography{bibliography.bib}
\end{document}